\documentclass[final,5p,times,twocolumn]{elsarticle}
\usepackage{bm}
\usepackage{amssymb,amsmath}

\journal{Physics Letters A}

\begin{document}

\begin{frontmatter}

\title{Relativistic analysis of the dielectric Einstein box:\\ Abraham, Minkowski and total energy-momentum tensors}
\author[udec]{Tom\'as Ramos}
\ead{tramos@udec.cl}
\author[udec]{Guillermo F. Rubilar}
\ead{grubilar@udec.cl}
\author[ucl]{Yuri N. Obukhov}
\ead{obukhov@math.ucl.ac.uk}
\address[udec]{Departamento de F\'isica, Universidad de Concepci\'on, Casilla 160-C, Concepci\'on, Chile}
\address[ucl]{Department of Mathematics and Institute of Origins,
University College London, Gower Street, London, WC1E 6BT, UK}
\begin{abstract}
\hspace*{0.4cm}We analyse the ``Einstein box'' thought experiment and the definition of the momentum 
of light inside matter. We stress the importance of the total energy-momentum tensor of the closed 
system (electromagnetic field plus material medium) and derive in detail the relativistic expressions 
for the Abraham and Minkowski momenta, together with the corresponding balance equations for 
an isotropic and homogeneous medium. We identify some assumptions hidden in the Einstein box 
argument, which make it weaker than it is usually recognized. In particular, we show that the 
Abraham momentum is not uniquely selected as the momentum of light in this case.
\end{abstract}


\end{frontmatter}

\section{Introduction}\label{intro}
The problem of the adequate definition of the momentum of light inside material media has 
surprisingly been discussed for more than 100 years since the original papers of Minkowski 
\cite{minkowski} and Abraham \cite{abraham}, and even so there is still some confusion or at 
least disagreement among authors. Minkowski proposed a non-symmetric energy-momentum 
tensor, which has the advantages that it can be directly derived from Maxwell's macroscopic 
equations and that it is related to the symmetries of the medium, when the latter is fixed 
(without dynamics). On the other hand, the Abraham tensor is symmetric, but cannot be derived 
from first principles \cite{obukhov1}. The Minkowski and Abraham momentum densities of a 
light pulse in a medium at rest, are defined as
\begin{align}
\bm{\pi}_M:=\bm{D}\times\bm{B},\qquad
\bm{\pi}_A:=\frac{1}{c^2}\,\bm{E}\times\bm{H}.\label{Adensity}
\end{align}

In the simple case of plane waves propagating within an isotropic and homogeneous medium 
at rest, the rival expressions (\ref{Adensity}) reduce to:
\begin{align}
\bm{\pi}_M=n\frac{{\cal U}}{c}\,\hat{\bm k},\qquad
\bm{\pi}_A=\frac{1}{n}\frac{\cal U}{c}\,\hat{\bm k}\label{pia0},
\end{align}
where $n$ is the refractive index of the medium, ${\cal U}$ is the 
energy density and $\hat{\bm k}$ is the propagation unit vector.
The difference between the very idealized predictions (\ref{pia0}) motivated 
the debate of determining which of the two definitions for the momentum density of light inside 
media is the correct one. In the literature we can find many theoretical discussions and a few 
experiments, which seem to favor one, both or neither of the two momenta for light discussed 
here. For a good and concise review of the Abraham-Minkowski controversy see the introduction 
of \cite{saldanha} and for a more detailed and historical review, not always free of confusion and 
contradictions, see \cite{pfeifer}-\cite{baxterloudon}.

The fundamentals of the controversy were understood in a formal manner more than 40 years ago, 
see \cite{penfieldhaus1}-\cite{israel}. Basically, it was recognized that only the total energy-momentum 
of the closed system consisting of electromagnetic field plus material medium has absolute 
physical meaning and that the Minkowski, Abraham and other expressions for the electromagnetic 
field simply correspond to different separations of the same total tensor. 
The expression for the total tensor will of course depend on the nature of the specific medium.

In the past 10 years, the discussion of the momentum of light inside media has become relevant 
again, specially due to the large number of more practical and ``quantum oriented'' works of 
Loudon, Barnett and collaborators, which analyze the radiation pressure based on the Lorentz 
force, \cite{padgettbarnettloudon}-\cite{hindsbarnett}. Mansuripur does a similar analysis, but 
using a modified force definition \cite{mansuripur1,mansuripur2}. In this group of papers, it is 
usually stated that the so-called ``Einstein box theories'', first proposed by Balazs \cite{balazs} 
as a modified Einstein thought experiment \cite{einstein}, uniquely select the Abraham 
momentum as the momentum of the field. Their main arguments are that the Minkowski 
momentum would predict a motion of the slab in the opposite direction to the incident pulse 
and that the Abraham momentum is the only one which simultaneously conserves the velocity 
of the center of energy, the total energy and the total momentum of the system. Here we explicitly 
show that using the Minkowski momentum with its adequate balance equations, i.e. with the 
total energy-momentum tensor, one arrives at the same results as with the Abraham momentum. 
Most recently, Barnett and Loudon in \cite{barnett2010,loudonbarnett3} reanalyzed the 
controversy and argued that both momenta are ``correct'', because both could be measured, 
but in different situations. They identify the Abraham momentum as the ``kinetic'' momentum 
of light and use Balazs's idea as their strongest argument to discard the Minkowski momentum 
in that situation. In other types of experiments, where the medium is at rest, it is 
claimed that the Minkowski momentum correctly describes the situation and therefore identify 
it as the ``canonical'' momentum of the field. In \cite{mansuripur3} it is also claimed that the 
controversy was solved, but in a different manner. One difference is that the author recognizes 
that the Einstein box argument does not uniquely determine the momentum of the field, but 
despite that he insists that this is the strongest available argument for the identification of 
light's momentum inside media and therefore considers the definition of the Abraham momentum 
for the field as an additional postulate of his theory. 

In this letter we make use of the general result given in \cite{obukhov1} for the total 
energy-momentum of the system consisting of electromagnetic field and an isotropic dielectric 
medium and derive, in a completely relativistic and self-consistent manner, the expressions for 
the Abraham and Minkowski momenta together with the corresponding balance equations 
for the case of a light pulse propagating inside a dielectric slab. This approach will allow us to 
identify some assumptions hidden in the Einstein box argument, which make it weaker than it 
is usually recognized.

\section{Relativistic model and the total energy-momentum tensor}

Suppose there is a dielectric slab of mass $M$ with homogeneous and isotropic electromagnetic 
properties, floating in space. In its rest frame, its index of refraction is $n$, its length is $L$ and it 
occupies a finite volume $V$. The slab is initially at rest, but a light pulse of total energy 
${\cal E}_0$ and finite volume $V_p\ll V$ strikes the slab from vacuum at normal incidence 
putting it in motion with a \emph{final} constant velocity $\bm{v}$. The slab is equipped with 
anti-reflection coatings so that the pulse can enter the slab at normal incidence without reflection 
and energy losses.

A fully relativistic model for the total energy-momentum tensor $T_{\mu}{}^{\nu}$ for a 
general linear, non-dissipative, non-dispersive and isotropic dielectric fluid with proper energy 
density $\rho$, pressure $p$, $4$-velocity field $u^{\mu}$, relative permittivity $\varepsilon$, 
relative permeability $\mu$ and particle number density $\nu$, interacting with the 
electromagnetic field $F_{\mu\nu}$ was first derived by Penfield and Haus in 1966 
\cite{penfieldhaus1} and is explicitly given in more modern form in the review \cite{obukhov1}. 
Following the same conventions as in \cite{obukhov1}, and neglecting gravitational as well as 
possible electro- and magnetostriction effects, and assuming negligible pressure $p\approx 0$,
we have
\begin{align}
T_{\mu}{}^{\nu}={}&\frac{\rho}{c^2}u_{\mu}u^{\nu}+\frac{1}{\mu\mu_0}\left(F_{\mu\sigma}F^{\sigma\nu}+\frac{1}{4}\delta_{\mu}^{\nu}F^{\sigma\lambda}F_{\sigma\lambda}\right)\nonumber\\
&+{}\frac{(n^2-1)}{\mu\mu_0 c^2}\left(F_{\mu\sigma}F^{\lambda\nu}u^{\sigma}u_{\lambda}+\frac{1}{2}\delta_{\mu}^{\nu}F_{\sigma\rho}F^{\sigma\lambda}u^{\rho}u_{\lambda}
\right.\nonumber\\
&\left.-\frac{1}{c^2}F^{\rho\sigma}F_{\rho\lambda}u_{\sigma}u^{\lambda}u_{\mu}
u^{\nu}\right).\label{tensortot2}
\end{align}
The total tensor is symmetric and satisfies the following energy-momentum balance equation,
\begin{align}
\partial_{\nu}T_{\mu}{}^{\nu}-F_{\mu\nu}J^{\nu}_{\rm ext}=0,\label{conservacioncero}
\end{align}
where the 4-vector $J_{\rm ext}^{\nu}$ describes the external charge and current densities 
which do not belong to the dielectric fluid. If $J_{\rm ext}^{\nu}=0$ energy-momentum tensor 
of the complete system is conserved and we have a closed system.

If we choose a volume $V'$ big enough so that it encloses the pulse and the slab until the pulse leaves the slab from the other 
side, then we can integrate the conservation equation and obtain that the total 4-momentum 
${\cal P}_{\mu}:=\left(E,-\bm{p}\right)$ of the whole system, defined as
\begin{align}
{\cal P}_{\mu}:=\int_{V'}T_{\mu}{}^{0}dV,\label{conservedpmu}
\end{align}
is a conserved, i.e. time independent, quantity. We will use this conservation of energy and momentum of the closed system to study the motion of the slab, when the pulse is propagating inside it (once the slab achieved a final constant velocity after a short deformation transient).

Because of the anti-reflection coatings, the light pulse can pass completely in the 
same incident direction (without reflection components), so that the problem can be treated just 
as a one-dimensional problem. Therefore the 4-velocity $u^{\mu}:=(\gamma,\gamma \bm{v})$ 
of the dielectric slab can be chosen to be
\begin{equation}
u^{\mu}=(\gamma,\gamma v,0,0),\label{cuadrivel}
\end{equation}
where $\gamma:=(1-\beta^2)^{-1/2}$ and $\beta:=v/c$ as usual.
Finally, we assume that the energy density distribution in the comoving frame is 
homogeneous:
\begin{equation}
\rho=\frac{Mc^2}{V}={\rm const}.\label{rhoo}
\end{equation}

\section{Energy-momentum tensors of the electromagnetic field}\label{abrahammink}

The total energy-momentum tensor (\ref{tensortot2}) can be split in different ways. For example, 
we can assign for the slab the energy-momentum tensor of a fluid without pressure (dust):
\begin{equation}
{\stackrel {\rm m} \Omega}{}_{\mu}{}^{\nu}:=\frac{\rho}{c^2}u_{\mu}u^{\nu},\label{abrmatter}
\end{equation}
and then light will be described by the Abraham energy-momentum tensor \cite{obukhov1}
\begin{align}
\Omega_{\mu}{}^{\nu}:={}&T_{\mu}{}^{\nu}-{\stackrel {\rm m} \Omega}{}_{\mu}{}^{\nu}\\
={}&\frac{1}{\mu\mu_0}\left(F_{\mu\sigma}F^{\sigma\nu}+\frac{1}{4}\delta_{\mu}^{\nu}F^{\sigma\lambda}F_{\sigma\lambda}\right)\nonumber\\
+{}&\frac{(n^2-1)}{\mu\mu_0 c^2}\left(F_{\mu\sigma}F^{\lambda\nu}u^{\sigma}u_{\lambda}+\frac{1}{2}\delta_{\mu}^{\nu}F_{\sigma\rho}F^{\sigma\lambda}u^{\rho}u_{\lambda}
\right.\nonumber\\
&\left.-\frac{1}{c^2}F^{\rho\sigma}F_{\rho\lambda}u_{\sigma}u^{\lambda}
u_{\mu}u^{\nu}\right).\label{qqqabr}
\end{align}
With this interpretation the total conserved 4-momentum of the system in (\ref{conservedpmu}) 
turns out to be ${\cal P}_{\mu}={\stackrel {\rm m,A} {\cal P}}{}_{\mu}
+{\stackrel {\rm A} {\cal P}}{}_{\mu}$,
where
\begin{align}
{\stackrel {\rm m,A} {\cal P}}{}_{\mu}:=\int_{V'}{\stackrel {\rm m} \Omega}{}_{\mu}{}^{0}dV,\qquad
{\stackrel {\rm A} {\cal P}}{}_{\mu}:=\int_{V'}\Omega_{\mu}{}^{0}dV.\label{pmufield}
\end{align}

On the other hand, we can consider that the electromagnetic energy and momentum content is 
described by the Minkowski energy-momentum tensor $\Theta_{\mu}{}^{\nu}$, whose definition 
follows directly from Maxwell's equations \cite{obukhov1},
\begin{align}
\Theta_{\mu}{}^{\nu}:={}&F_{\mu\sigma}H^{\sigma\nu}+\frac{1}{4}\delta_{\mu}^{\nu}F_{\sigma\lambda}H^{\sigma\lambda}\\
={}&\frac{1}{\mu\mu_0}\left(F_{\mu\sigma}F^{\sigma\nu}+\frac{1}{4}\delta_{\mu}^{\nu}F_{\sigma\lambda}F^{\sigma\lambda}\right)
+\frac{(n^2-1)}{\mu\mu_0 c^2}\left(F_{\mu\lambda}F^{\lambda\rho}u_{\rho}u^{\nu}\right.\nonumber\\
{}&\left.+F_{\mu\sigma}F^{\lambda\nu}u^{\sigma}u_{\lambda}+\frac{1}{2}\delta_{\mu}^{\nu}F_{\sigma\rho}F^{\sigma\lambda}u^{\rho}u_{\lambda}\right).\label{explicitmink}
\end{align}
This tensor can be obtained by adding the term
\begin{align}
Q_{\mu}{}^{\nu}:=\frac{(n^2-1)}{\mu\mu_0 c^2}\left[F_{\mu\lambda}F^{\lambda\rho}u_{\rho}u^{\nu}+\frac{1}{c^2}F^{\rho\sigma}F_{\rho\lambda}u_{\sigma}u^{\lambda}u_{\mu}u^{\nu}\right]\label{qqqq}
\end{align}
to the Abraham tensor, so that
\begin{equation}
\Theta_{\mu}{}^{\nu}=\Omega_{\mu}{}^{\nu}+Q_{\mu}{}^{\nu}.\label{def1}
\end{equation}
Consequently, the total energy-momentum tensor can be written as $T_{\mu}{}^{\nu}
={\stackrel {\rm m} \Theta}{}_{\mu}{}^{\nu} + \Theta_{\mu}{}^{\nu}$, where
\begin{align}
{\stackrel {\rm m} \Theta}{}_{\mu}{}^{\nu} :={}&{\stackrel {\rm m} \Omega}{}_{\mu}{}^{\nu}-Q_{\mu}{}^{\nu}\label{def2}\\
={}&\frac{\rho}{c^2}u_{\mu}u^{\nu}-\frac{(n^2-1)}{\mu\mu_0 c^2}\left[F_{\mu\lambda}F^{\lambda\rho}u_{\rho}u^{\nu}+\frac{1}{c^2}F^{\rho\sigma}F_{\rho\lambda}u_{\sigma}u^{\lambda}u_{\mu}u^{\nu}\right],
\end{align}
is the Minkowski energy-momentum tensor for matter. Finally, with this interpretation, the total 
conserved 4-momentum ${\cal P}_{\mu}$ can be also expressed as $
{\cal P}_{\mu}={\stackrel {\rm m,M} {\cal P}}{}_{\mu} 
+ {\stackrel {\rm M} {\cal P}}{}_{\mu}$, where
\begin{align}
{\stackrel {\rm m,M}{\cal P}}_\mu:=\int_{V'} {\stackrel {\rm m} \Theta}{}_{\mu}{}^{0}dV,\qquad
{\stackrel {\rm M}{\cal P}}_\mu:=\int_{V'} \Theta_{\mu}{}^{0}dV,
\end{align}
are the Minkowski 4-momenta for matter and electromagnetic field, respectively.
\section{Explicit calculation with the Abraham tensor}
\subsection{Abraham energy and momentum for the slab}

We derive first the Abraham tensor, which is more compact in this case. 
If we substitute (\ref{cuadrivel}) into (\ref{pmufield}a) and use the identification 
${\stackrel {\rm m,A} {\cal P}}_{\mu} = ({\stackrel {\rm m,A} E}, -{\stackrel {\rm m,A} {\bm{p}}})$, then the Abraham energy and momentum for the slab read
\begin{align}
{\stackrel {\rm m,A} E}={}&\int_{V'} \rho\,{\frac {u_{0}u^{0}}{c^2}}dV,\\
{\stackrel {\rm m,A} p}{}_{a}={}&-\int_{V'} \rho\,{\frac {u_{a}u^{0}}{c^2}}dV,\quad a=1,2,3.
\end{align}
Therefore, by explicitly calculating the integrals, we obtain
\begin{align}
{\stackrel {\rm m,A} E}=\rho\gamma^2 V_{v},\qquad
{\stackrel {\rm m,A} p}_a=-\frac{1}{c^2}g_{ab}v^b\rho\gamma^2 V_{v},
\end{align}
where $V_v$ is the volume of the slab in the reference frame where it moves with velocity 
$\bm{v}=v\,\hat{\bm x}$. Using the relation $V_v=V/\gamma$ and expression 
(\ref{rhoo}), we have
\begin{align}
{\stackrel {\rm m,A} E}=\gamma Mc^2,\qquad
{\stackrel {\rm m,A} {\bm{p}}}=\gamma M v\,\hat{\bm x}.\label{pmeda}
\end{align}
These results (\ref{pmeda}) are the usual expressions for the 
(relativistic) energy and momentum of a body of mass $M$ moving with velocity 
$v\hat{\bm x}$, which is not surprising because of our choice (\ref{abrmatter}) for the 
energy and momentum of the medium. The relation between momentum and energy is 
also the usual one for a relativistic massive particle:
\begin{equation}
{\stackrel {\rm m,A} {\bm{p}}}=v\ \frac{{\stackrel {\rm m,A} E}}{c^2}\hat{\bm x}.\label{pmv}
\end{equation}

\subsection{Abraham energy and momentum for the light pulse}
Using (\ref{cuadrivel}), (\ref{qqqabr}), (\ref{pmufield}b) and the identifications of the 
components of $F_{\mu\nu}$ in \cite{obukhov1}, we can explicitly compute the energy 
and momentum associated to the Abraham tensor for the electromagnetic field in terms 
of $\bm{E}$, $\bm{B}$ and $\bm{v}$:
\begin{align}
{\stackrel {\rm A} E}={}&\frac{1}{2\mu\mu_0}\int_{V'}(\bm{E}^2/c^2+\bm{B}^2)dV\nonumber\\
{}&-\frac{(n^2-1)}{2\mu\mu_0 c^2}\int_{V'}\left\{\gamma^2(2\gamma^2-1)
\left[(\bm{E}\cdot\bm{v})^2/c^2+(\bm{B}\cdot\bm{v})^2-\bm{E}^2\right.\right.\nonumber\\
{}& -\left.\left.\bm{v}^2\bm{B}^2-2\bm{E}\cdot(\bm{v}\times\bm{B})\right]-2\gamma^2(\bm{E}\cdot\bm{v})^2/c^2\ \right\}dV,\label{eabraham2}\\
{\stackrel {\rm A} {\bm{p}}}={}&\frac{1}{\mu\mu_0 c^2}\int_{V'}\bm{E}\times\bm{B}\ dV\nonumber\\
{}& -\frac{(n^2-1)}{\mu\mu_0 c^4}\int_{V'}\left\{\gamma^2 (\bm{E}\cdot\bm{v})(\bm{E}+\bm{v}\times\bm{B})
+\gamma^4\bm{v}\left[(\bm{E}\cdot\bm{v})^2/c^2\right.\right.\nonumber\\
{}& +\left.\left. (\bm{B}\cdot\bm{v})^2-\bm{E}^2-\bm{v}^2\bm{B}^2-2\bm{E}\cdot(\bm{v}\times\bm{B})\right]\right\}\ dV.\label{paivec}
\end{align}
From (\ref{eabraham2}) and (\ref{paivec}) we see that to zeroth order in 
$\bm{v}$ they reduce to the well known expressions for a linear, isotropic and 
homogeneous medium at rest,
\begin{align}
{\stackrel {\rm A} E}_{(0)}=\frac{1}{2}\int_{V'}(\bm{E}\cdot\bm{D}+\bm{B}\cdot\bm{H})\ dV,\qquad
{\stackrel {\rm A} {\bm{p}}}_{(0)}=\frac{1}{c^2}\,\int_{V'}\bm{E}\times\bm{H}\ dV,\label{panr0}
\end{align}
where we used the constitutive relations in the medium at rest 
$\bm{D}=\varepsilon\varepsilon_0\bm{E}$ and $\bm{H}=\bm{B}/\mu\mu_0$.

We consider the special simple case in which the electromagnetic pulse can be approximated by 
a ``cut'' plane wave of finite volume $V_p\ll V$, i.e. much less than the volume of the slab, but 
big enough so that the continuum approximation for the slab is valid, propagating in the 
direction $\hat{\bm{x}}$ of the slab's motion. Solving the macroscopic Maxwell equations 
inside the medium for light with any polarization (see \ref{appa} for more details), 
we can write,
\begin{align}
\bm{B}(x,t)={}&\frac{1}{v_{\beta}}\hat{\bm{x}}\times\bm{E}\label{bfield},
\end{align}
where $v_{\beta}$ is the phase velocity of light inside the moving medium, given by
\begin{equation}
v_{\beta}:=c\frac{(1+n\beta)}{(n+\beta)}.\label{phasevelocity}
\end{equation}
Since in our one-dimensional case $\bm{v}\perp\bm{E}$ and $\bm{v}\perp\bm{B}$, the 
expressions (\ref{eabraham2}) and (\ref{paivec}) reduce to:
\begin{align}
{\stackrel {\rm A} E}={}&\frac{1}{2\mu\mu_0c^2}\int_{V'}(\bm{E}^2+c^2\bm{B}^2)dV\nonumber\\
{}&+\frac{(n^2-1)}{2\mu\mu_0c^2}\int_{V'}\gamma^2(2\gamma^2-1)|\bm{E}+\bm{v}\times\bm{B}|^2\ dV,\label{ea3}\\
{\stackrel {\rm A} {\bm{p}}}={}&\frac{1}{\mu\mu_0 c^2}\,\int_{V'}\bm{E}\times\bm{B}\,dV
+\frac{(n^2-1)}{\mu\mu_0 c^4}\,\int_{V'}\gamma^4\bm{v}|\bm{E}+\bm{v}\times\bm{B}|^2 dV.\label{pa3}
\end{align}
If we insert (\ref{bfield}) and (\ref{phasevelocity}) into (\ref{ea3}) and (\ref{pa3}), we obtain more compact expressions for ${\stackrel {\rm A} E}$ and ${\stackrel {\rm A} {\bm{p}}}$, just in terms of $\bm{E}^2$:
\begin{align}
{\stackrel {\rm A} E}={}&\frac{1}{\mu\mu_0c^2}\int_{V'}\frac{n(n+2\beta+n\beta^2)}{(1+n\beta)^2}\bm{E}^2\ dV,\label{ea5}\\
{\stackrel {\rm A} {\bm{p}}}={}&\frac{1}{\mu\mu_0 c^3}\hat{\bm x}\int_{V'}\frac{n(1+2n\beta+\beta^2)}{(1+n\beta)^2}\bm{E}^2\ dV.\label{pa5}
\end{align}
When the pulse is fully inside the slab, we can integrate over the volume $V_p$ of the pulse and the factors with $n$ and $\beta$ will go out the integral. Therefore, we can relate the Abraham momentum and Abraham energy of the pulse inside the medium by
\begin{equation}
{\stackrel {\rm A} {\bm{p}}}=\frac{(1+2n\beta+\beta^2)}{(n+2\beta+n\beta^2)}\frac{{\stackrel {\rm A} E}}{c}\,\hat{\bm x},\label{proper}
\end{equation}
which is an important result that we will use in the next subsection. It is worthwhile to notice that 
(\ref{proper}) is valid for any polarization of the ``cut'' plane-wave pulse, but not for a general 
pulse form since, if we compare (\ref{ea3}) with (\ref{pa3}), we see that 
${\stackrel {\rm A} {\bm{p}}}$ and ${\stackrel {\rm A} E}$ are not proportional in general.
As a consistency test, it can be checked that the same result (\ref{proper}) can be obtained if 
we apply a boost to the well-known Abraham expression (\ref{pia0}b) valid in the rest frame of 
the medium.
\subsection{Conservation of the center of energy velocity}

In \cite{barnett2010}, Barnett revitalized the argument of Balazs \cite{balazs} which states 
that the conservation of the center of energy velocity, in addition to the conservation of 
momentum, uniquely selects the momentum of light inside the slab to be the one of Abraham. 
We will now examine these arguments in more detail. We can check from (\ref{proper}) that 
when the medium is at rest, we get the typical value of the Abraham momentum, see (\ref{pia0}b),  
${\stackrel {\rm A} {\bm{p}}}=(\stackrel {\rm A}{E}/nc)\hat{\bm x}$, 
which we can write in terms of the phase velocity of light in that reference frame $v_0:=c/n$ as
\begin{equation}
{\stackrel {\rm A} {\bm{p}}}=v_{0}\frac{{\stackrel {\rm A} E}}{c^2}\hat{\bm x},\label{pbarnett}
\end{equation}
i.e. as if it was a particle of ``moving mass'' ${\stackrel {\rm A} M}:={\stackrel {\rm A} E}/c^2$ and velocity $\bm{v}_0=(c/n)\hat{\bm x}$, in a way similar to (\ref{pmv}).
In \cite{barnett2010,loudonbarnett3} the explicit definition of the conserved ``center of energy velocity'' of the system $v_{\rm CM}$ is not shown, however, from special relativity we know that it is related to the total energy $E$ and the total momentum $\bm{p}$ of the system, by
\begin{align}
\bm{v}_{\rm CE}=\frac{c^2}{E}\bm{p}.\label{defvce}
\end{align}
Now, in order to reproduce Barnett's argument in detail, we add the momentum of light in the form (\ref{pbarnett}) to the momentum of the slab in (\ref{pmv}) and use the total energy and momentum conservation to get
\begin{equation}
v'_{\rm CE}=\frac{{\stackrel {\rm A} E}\,(c/n)+{\stackrel {\rm m,A} E}v}{{\stackrel {\rm A} E}+{\stackrel {\rm m,A} E}}.
\end{equation}

Strictly speaking this argument is \textit{incorrect}, because $v'_{\rm CE}$ is not a conserved quantity. The expression (\ref{pbarnett}) for the electromagnetic field is only valid when the slab is at rest, but the final velocity of the medium is not zero and the choice of the velocity $v_0=c/n$ for the light pulse is inappropriate. If we want to write the expression of the momentum of the pulse as if it were a particle with energy ${\stackrel {\rm A} E}$, then as can be seen from (\ref{proper}), the proper ``particle'' velocity $v_{\rm p}$ should be defined as
\begin{equation}
v_{\rm p}:=c\frac{(1+2n\beta+\beta^2)}{(n+2\beta+n\beta^2)},\label{vpp}
\end{equation}
and hence the correct total momentum of the system would have the form
$\bm{p}=(\stackrel {\rm A}{E}/c^2)v_{\rm p}\hat{\bm x}+(\stackrel {\rm m,A}{E}/c^2)v\hat{\bm x}$.
Together with the total energy of the system $E={\stackrel {\rm m,A} E}+{\stackrel {\rm A} E}$, which is also conserved, the velocity of the center of energy $\bm{v}_{CE}$ given by
\begin{align}
\bm{v}_{CE}={}&\frac{{\stackrel {\rm A} E}\,v_{\rm p}+{\stackrel {\rm m,A} E}v}{{\stackrel {\rm A} E}+{\stackrel {\rm m,A} E}}\,\hat{\bm x}\label{defvce2},
\end{align}
turns out to be a conserved quantity indeed.
In general, the center of energy velocity $\bm{v}_{CE}$ in (\ref{defvce}) is always a conserved quantity in a closed system. Since the total energy-momentum tensor (\ref{tensortot2}) of the closed system is conserved and symmetric, all the components of the total angular 4-momentum tensor $L_{\rho\sigma}:=\int_{V_p}(x_{\rho}T_{\sigma}{}^{0}-x_{\sigma}T_{\rho}{}^{0})$ are conserved as well, including the components $0a$ associated to the conservation of the velocity of the center of energy. It is important to remark that this conservation holds independently of the choice of the Abraham or the Minkowski momentum to describe the electromagnetic field, because it depends on the total quantities $\bm{p}$ and $E$.

Although formally incorrect, in practice the naive expression $v_0=c/n$ yields a very good approximation for the particle velocity of the pulse. As we will see in section \ref{cuatrocinco}, $\beta\sim10^{-15}$ in a standard case and $\beta\sim10^{-9}$ in an extreme case, so if we expand (\ref{vpp}),
\begin{align}
v_{\rm p}=\frac{c}{n}+2\frac{c(n^2-1)}{n^2}\beta+O(\beta^2).\label{vpapprox}
\end{align}
we see that $v_{\rm p}\approx v_0=c/n$ is an extremely accurate approximation indeed. It is also remarkable, that the correct velocity that should enter (\ref{defvce2}) is also different from the relativistic phase velocity $v_{\beta}$ in (\ref{phasevelocity}), as one could naively expect as a generalization of $v_0$. The non-relativistic expansion of $v_{\beta}$ reads,
\begin{align}
v_{\beta}=\frac{c}{n}+\frac{c(n^2-1)}{n^2}\beta+O(\beta^2),\label{vbetaapprox}
\end{align}
differing from (\ref{vpapprox}) by a factor $2$ in the term of first order in $\beta$. This term in (\ref{vbetaapprox}) has been measured with great accuracy in the Fizeau experiment. See, for instance \cite{brevik}, page 187.

\subsection{Conservation equations and solution of the slab motion}\label{cuatrocinco}
Since we already know the explicit forms of the energy and momentum of the complete system in the Abraham separation, we can use the two conservation equations to solve the problem of the motion of the slab in terms of the system's parameters. We will consider two states of the system, first when the pulse is traveling with total energy ${\cal E}_0$ in vacuum and the slab of mass $M$ and refraction index $n$ is at rest, and finally when the electromagnetic pulse is completely inside the slab after it already reached a final constant velocity $\bm{v}=c\beta\hat{\bm x}$. Therefore, the energy conservation equation reads
\begin{align}
{\stackrel {\rm m,A} E}{}^{\rm(out)}+{\stackrel {\rm A} E}{}^{\rm(out)}={}&{\stackrel {\rm m,A} E}{}^{\rm(in)}+{\stackrel {\rm A} E}{}^{\rm(in)},\\
Mc^2+{\cal E}_0={}&\gamma Mc^2+{\stackrel {\rm A} E},\label{eenergy}
\end{align}
and the total momentum conservation in the propagation direction $\hat{\bm x}$ is given by
\begin{align}
{\stackrel {\rm m,A} {\bm{p}}}{}^{\rm(out)}+{\stackrel {\rm A} {\bm{p}}}{}^{\rm(out)}={}&{\stackrel {\rm m,A} {\bm{p}}}{}^{\rm(in)}+{\stackrel {\rm A} {\bm{p}}}{}^{\rm(in)},\\
0\ +\frac{{\cal E}_0}{c}={}&\gamma Mc\beta+\frac{(1+2n\beta+\beta^2)}{(n+2\beta+n\beta^2)}\frac{{\stackrel {\rm A} E}}{c}.\label{emomentum}
\end{align}
Equations (\ref{eenergy}) and (\ref{emomentum}) constitute a system of two equations for two unknowns $\beta$ and ${\stackrel {\rm A} E}$, which we can solve in terms of the system parameters ${\cal E}_0$, $M$ and $n$.
From (\ref{eenergy}) we can find an expression for ${\stackrel {\rm A} E}$ in terms of $\beta$, which reads
\begin{equation}
{\stackrel {\rm A} E}={\cal E}_0+Mc^2(1-\gamma).\label{eaalone}
\end{equation}
This last equation already determines that the motion of the slab will be non-relativistic in most practical situations. Let us consider the extreme case when all the energy of the light pulse in vacuum is transformed in kinetic energy of the slab. Then in (\ref{eaalone}), ${\stackrel {\rm A} E}=0$, and we can determine $\gamma_{\rm max}$ as,
\begin{align}
\gamma_{\rm max}=1+q,\label{max}
\end{align}
where we have defined the dimensionless parameter $q$ by
\begin{equation}
q:=\frac{{\cal E}_0}{Mc^2}.\label{defffq}
\end{equation}
The parameter $q$ (together with $n$) determines the motion of the slab. In practice $q$ is extremely small, as we shall see at the end of the section, of the order $q\sim 10^{-9}$ or less, so from (\ref{max}) we see that $\gamma_{\rm max}$ will be very close to unity and therefore $\beta_{\rm max}$ will be at most of the order $\beta_{\rm max}\sim\sqrt{2q}\sim 10^{-4}\ll 1$, resulting in a non-relativistic motion of the slab.
Even though we know that the motion will be non-relativistic, we will first present the full equation for $\beta$, without any further approximation. Inserting (\ref{eaalone}) in (\ref{emomentum}), we get a fourth order polynomial equation for $\beta$ in terms of the parameters $q$ and $n$:
\begin{align}
{}&[(1+q-nq)^2+n^2]\beta^4+[4(1+q-nq)(n-q+nq)+2n]\beta^3\nonumber\\
&+[2(1+q-nq)^2+4(n-q+nq)^2+1-n^2]\beta^2\nonumber\\
{}&+[4(1+q-nq)(n-q+nq)-2n]\beta+[(1+q-nq)^2-1]=0.\label{eq2}
\end{align}
This equation can be solved analytically, but the expression of the solution is large and not instructive. Since we already know that in practice $q\ll1$ and $\beta\ll1$, it is interesting to search for a more tractable approximated solution for $\beta$. Therefore, if we keep only the first order terms in (\ref{eq2}), we get the well known solution of Balazs, Barnett, Mansuripur and many others, for the non-relativistic velocity of the dielectric slab
\begin{equation}
\beta\approx\frac{1}{n}(n-1)q,\label{eq6}
\end{equation}
or, in more familiar terms,
\begin{equation}
v\approx\frac{(n-1)}{n}\frac{{\cal E}_0}{Mc}>0,\label{eq7}
\end{equation}
which means that the slab will move in the same direction of the electromagnetic pulse, while 
the pulse is propagating inside it. If we continue in the non-relativistic limit, the light pulse 
will spend a time interval $\Delta t\approx nL/c$ inside the slab and therefore its net 
displacement $\Delta x$ will be, as usual, $\Delta x\approx(n-1)L{{\cal E}_0}/{Mc^2}>0$.
Additionally, one can also find a solution of (\ref{eq2}) which is exact to second order in $q$:
\begin{align}
\beta(n,q)={}&\frac{(n-1)}{n}q-\frac{(4n^3-5n^2-2n+3)}{2n^3}q^2+O(q^3),\label{correction}
\end{align}
where the second term can be considered as the first ``relativistic correction'' for $\beta$. From (\ref{correction}) we can estimate the error that we make by using just the first order non-relativistic approximation (\ref{eq6}). Suppose that the slab is made of glass with $n=1.5$. If it has a mass of the order $M\sim 100 g$ and if the light source is a good pulsed laser with energy ${\cal E}_0\sim 1 J$, then the parameter $q$ will be typically of the order $q\sim10^{-15}$ and therefore from (\ref{eq7}) we see that $v\sim 0.1\ \mu m/s$. In this case, the difference between the second order solution of (\ref{correction}) and the non-relativistic solution (\ref{eq6}) is of the order $\Delta\beta\sim 10^{-31}$ and the relative error of using (\ref{eq6}) is $\sim100q\sim10^{-13}\%$. In the extreme case of the most powerful lasers available ${\cal E}_0\sim 1 kJ$ and a very small dielectric slab with mass $M\sim 10 g$, we would in theory be able to achieve a $q$ parameter of the order $q\sim10^{-9}$ and a final slab velocity of order $v\sim10\ cm/s$. In this case, $\Delta\beta\sim 10^{-19}$ and the relative error is $\sim100q\sim10^{-7}\%$ and hence we see that the well-known non-relativistic solution (\ref{eq7}) is extremely accurate for all practical purposes.

\section{Using the Minkowski tensor}\label{secMink}
\subsection{Minkowski energy and momentum expressions}

In \cite{brevik,milonniboyd,padgettbarnettloudon,padgett,balazs,barnett2010,bradshawboydmilonni} and other papers, it is argued that the Minkowski momentum fails to describe this slab experiment by predicting the motion of the slab in the opposite direction of the incident electromagnetic pulse. However, the mistake is to consider the same balance equations that are valid for the Abraham momentum, while using the Minkowski momentum for the electromagnetic pulse, thereby tacitly assigning an incorrect energy and momentum to matter in the Minkowski picture.
Jones \cite{jones78} noticed this deficiency and suggested that the correct 
momentum of matter should include the ``forward bodily impulse'' the nature of
which he was unable to describe. The explanation is simple, though: one needs
to use the {\it canonical momentum} of matter which, combined with the canonical
momentum of the electromagnetic field, is conserved in the Minkowski picture.

As we mentioned in section \ref{abrahammink}, we can formally compute the correct Minkowski quantities from the Abraham expressions for the energy and momentum by adding the proper components of $Q_{\mu}{}^{\nu}$. Evaluating the expression for $Q_{\mu}{}^{0}$ in (\ref{qqqq}), we get
\begin{equation}
Q_{\mu}{}^{0}=\frac{(n^2-1)}{\mu\mu_0c^2}\frac{n}{(1+n\beta)^2}\bm{E}^2\left(\beta,-1,0,0\right).{\label{expq}}
\end{equation}
Then, the Minkowski energy of the pulse is of the form
\begin{align}
{\stackrel {\rm M} E} = {}&{\stackrel {\rm A} E}+\int_{V'}Q_{0}{}^{0}dV,\\
={}&\frac{n}{\mu\mu_0c^2}\frac{c}{v_{\beta}}\int_{V_p}\bm{E}^2\ dV.\label{menergy}
\end{align}
Using (\ref{ea3}) integrated over $V_p$, we can relate ${\stackrel {\rm M}E}$ to the Abraham energy of the light pulse by
\begin{equation}
{\stackrel {\rm A} E}=\frac{(n+2\beta+n\beta^2)}{(1+n\beta)(n+\beta)}{\stackrel {\rm M} E}.\label{relationam}
\end{equation}
These energies coincide, for a given $n\neq 1$, only in the case $\beta=0$, i.e. in the rest frame of the medium.
In the same way, starting from the definitions (\ref{def1}) and (\ref{def2}), and using the expressions (\ref{expq}) and (\ref{menergy}), we can determine all the other Minkowski quantities in terms of ${\stackrel {\rm M} E}$:
\begin{align}
{\stackrel {\rm M} {\bm{p}}}={}&\frac{{\stackrel {\rm M} E}}{v_{\beta}}\,\hat{\bm x},\label{mp2}\\
{\stackrel {\rm m,M} E}={}&\gamma Mc^2-\frac{(n^2-1)\beta}{(1+n\beta)(n+\beta)}{\stackrel {\rm M} E},\label{mmede}\\
{\stackrel {\rm m,M} {\bm{p}}}={}&\left[\gamma M c\beta-\frac{(n^2-1)}{(1+n\beta)(n+\beta)}\frac{{\stackrel {\rm M} E}}{c}\right]\hat{\bm x}.\label{mmedp}
\end{align}
The last term in the {\it canonical momentum} of the matter (\ref{mmedp}) 
accounts for the ``forward bodily impulse'' of Jones \cite{jones78}. 

\subsection{Defining a Minkowski velocity}
With the results (\ref{menergy}), (\ref{mp2}), (\ref{mmede}) and (\ref{mmedp}) we can express the center of energy velocity in the Minkowski picture as
\begin{align}
v_{CE}={}&\frac{c^2}{E}\left[\gamma M c\beta-\frac{(n^2-1)}{(1+n\beta)(n+\beta)}\frac{{\stackrel {\rm M} E}}{c}+\frac{{\stackrel {\rm M} E}}{v_{\beta}}\right]\\
={}&\frac{c^2}{E}\left[v\ \frac{\gamma M c^2}{c^2}+\frac{c(1+2n\beta+\beta^2)}{(1+n\beta)(n+\beta)}\,\frac{{\stackrel {\rm M} E}}{c^2}\right]\\
={}&\frac{v\ (\gamma M c^2)+v_{\rm M}\ {\stackrel {\rm M} E}}{E},
\end{align}
where we defined the Minkowski velocity $v_{\rm M}$ of the field as
\begin{align}
v_{\rm M}:=c\ \frac{(1+2n\beta+\beta^2)}{(1+n\beta)(n+\beta)}.\label{vM}
\end{align}
To get the velocity (\ref{vM}), we shifted terms from ${\stackrel {\rm m,M} p}$ to ${\stackrel {\rm M} p}$ and therefore it does not satisfy the prescription of a ``particle velocity'',
\begin{align}
\bm{v}_{\rm em}=c^2\frac{\bm{p}_{\rm em}}{E_{\rm em}},\qquad \bm{v}_{\rm mat}=c^2\frac{\bm{p}_{\rm mat}}{E_{\rm mat}}, \label{particlepres}
\end{align}
as for the case of Abraham light and matter velocities (\ref{pmv}) and (\ref{vpp}). In fact, $v_M$ in (\ref{vM}) is a very artificial velocity that one would have to associate to the field with the Minkowski energy, in order to keep the velocity $v$ for the block, and still obtain the correct (relativistic) center of energy velocity (\ref{defvce2}). If we expand (\ref{vM}) in powers of $\beta$ we see that $v_{\rm M}$ is also similar to all the other previously defined velocities,
\begin{align}
v_{\rm p,M}=\frac{c}{n}+\frac{c(n^2-1)}{n^2}\beta+O(\beta^2),
\end{align}
and hence equals $c/n$ in the non-relativistic limit to zeroth order in $\beta$, as well as the Abraham particle velocity.
We could also define a Minkowski ``particle'' velocity following the prescription (\ref{particlepres}). Therefore, using (\ref{mp2}), (\ref{mmede}) and (\ref{mmedp}), we have
\begin{align}
v_{\rm p,M}:={}&\frac{c^2{\stackrel {\rm M} p}}{{\stackrel {\rm M} E}}=\frac{c^2}{v_{\beta}}=c\frac{(n+\beta)}{(1+n\beta)},\label{vpM}\\
v_{\rm mat,M}:={}&\frac{c^2{\stackrel {\rm m,M} p}}{{\stackrel {\rm m,M} E}}=c\frac{\left[\gamma Mc^2(1+n\beta)(n+\beta)\beta-(n^2-1){\stackrel {\rm M} E}\right]}{\left[\gamma Mc^2(1+n\beta)(n+\beta)-(n^2-1)\beta {\stackrel {\rm M} E}\right]}.\label{vmatM}
\end{align}
With (\ref{vpM}) and (\ref{vmatM}) the expression for the velocity of ``center of energy" naturally assumes the form $v_{CE}=\left(\sum_i v_i\cdot E_i\right)/\left(\sum_i E_i\right)$, because it depends only on the total quantities, \textit{but neither (\ref{vpM}) nor (\ref{vmatM}) coincide with a velocity of the system which one can easily identify and interpret} (like the velocity of the block $v$ or the phase velocity of the field $v_{\beta}$, for example). Indeed, $v_{\rm mat,M}$ in (\ref{vmatM}) depends on the energy of the pulse, which is very counter-intuitive. When there is no light pulse, i.e. ${\stackrel {\rm M} E}=0$ and $v_{\rm mat,M}$ reduces to the velocity of the block $v$.
\subsection{Minkowski balance equations}

Since we already know all the explicit expressions for the Minkowski momentum and energy of the field and slab, we can use them to write the balance equations and correctly solve for the slab velocity also with the Minkowski formulation. From (\ref{mp2}), (\ref{mmede}) and (\ref{mmedp}), we have
\begin{align}
Mc^2+{\cal E}_0={}&\left[\gamma Mc^2-\frac{(n^2-1)\beta}{(1+n\beta)(n+\beta)}{\stackrel {\rm M} E}\right]+{\stackrel {\rm M} E},\label{meq1}\\
\frac{{\cal E}_0}{c}={}&\left[\gamma M c\beta-\frac{(n^2-1)}{(1+n\beta)(n+\beta)}\frac{{\stackrel {\rm M} E}}{c}\right]+\frac{(n+\beta)}{(1+n\beta)}\frac{{\stackrel {\rm M} E}}{c}.\label{meq2}
\end{align}
Taking (\ref{meq1}) and dividing it by $Mc^2$, we get
\begin{align}
q_M=\frac{(1+n\beta)(n+\beta)}{(n+2\beta+ n\beta^2)}(q+1-\gamma),\label{meq3}
\end{align}
where $q_M:=\frac{{\stackrel {\rm M} E}}{Mc^2}$, following the definition (\ref{defffq}).
Then, if we divide (\ref{meq2}) by $Mc$ and use (\ref{meq3}), we obtain after some algebra the same fourth order equation for $\beta$ in (\ref{eq2}), which we obtained with the Abraham formulation. Therefore in the the Minkowski picture, we obtain the same solution $\beta=\beta(n,q)$ for the motion of the slab. The authors who claim that the Minkowski momentum is unable to describe the slab plus light pulse system use the equations (\ref{meq1}) and (\ref{meq2}), but without the second terms inside the bracket on the r.h.s. and hence they use incorrect balance equations.

\section{Conclusions}
The fundamental equations, which govern the interactions and motion within the total system of electromagnetic field plus a material medium, are the macroscopic Maxwell equations, the constitutive relations and the hydrodynamic equations for the fluid. The energy and momentum balance equations for the total system have a clear physical meaning and we can use them, as here explicitly shown, to determine the dynamics of the system. The Abraham and Minkowski momenta can be understood as different separations of the total momentum, a choice of which does not affect the physical predictions.

When we use the total energy-momentum tensor, the conservation of momentum and of the velocity of ``center of energy" is always satisfied for any specific separation, because it involves only the total quantities. However, when we use the ``Abraham separation", we assign the energy-momentum tensor of a perfect fluid to the material subsystem, as if it were in isolation. As a result, the definition (\ref{particlepres}) of the Abraham velocity of matter coincides with the velocity of the block and it is the only separation in which this happens. This fact explains why the Abraham tensor is relevant for the case of an homogeneous and isotropic medium, and for the Einstein box theories in particular. Since in this picture we can consistently interpret the block as a particle, the remaining term can be naturally interpreted as the momentum of a ``light particle" with velocity $v_p$ given in (\ref{vpp}), which in the non-relativistic approximation (very accurate for these cases, as we have demonstrated) coincides with the phase velocity of light when the medium is at rest $v_0=c/n$. In other words, with the ``Abraham separation'' the property of inertia of energy is not only satisfied in the total system, as special relativity requires, but also in each subsystem separately. This was already noticed by Brevik in 1979, see \cite{brevik}, page 192.

In our opinion, these are the best arguments which support the usefulness of the interpretation of the Abraham momentum as the ``kinetic momentum" of the field when the momentum of light is introduced in the usual non-relativistic ``$mv$" form,
\begin{equation}
{\stackrel{{\rm A}}p} := {\stackrel{{\rm A}}M}v_0 = \frac{\stackrel{\rm A}E}{c^2}\frac{c}{n} = \frac{1}{n}\frac{\stackrel{\rm A}E}{c}.\label{mvvv}
\end{equation}
At the same time, in spite of the fact that the Abraham choice is simpler than Minkowski's one for the case of the block, our analysis in section \ref{secMink} clearly demonstrates that the Minkowski definition is also perfectly consistent for the Einstein box experiment, contrary to what is sometimes claimed \cite{brevik,milonniboyd,padgettbarnettloudon,padgett,balazs,barnett2010,bradshawboydmilonni},
provided one considers the correct Minkowski expressions for the energy and momentum of matter.  For a complementary discussion considering other explicit field configurations, see \cite{brevik2011}. Again, only the total quantities are relevant for the description of the system.

In the nonrelativistic discussion of the balance equations for the slab, see for instance \cite{balazs,barnett2010,loudonbarnett3,mansuripur3}, the Abraham momentum is selected as a consequence of treating the contribution of the light pulse to the velocity of the center of energy as if it were a particle moving with the phase velocity of the wave. We want to stress that this choice is not justified from the point of view of field theory. Additionally, our fully relativistic analysis shows that this assumption would only be consistent with the global conservation laws of the total system if one introduces suitable (ad hoc) ``particle velocities'' for the pulse, both in the Minkowski and Abraham pictures. However, these velocities do not correspond in general to any well defined velocity in the system. In particular, they do not coincide with the phase velocity of the wave in the moving medium, which is only true to zeroth order in the final slab velocity.

The Abraham tensor is useful for isotropic media \cite{obukhov1}, in the sense that if the block is described by a dust energy-momentum tensor, all other terms are contained in the Abraham tensor. However, this may not be the case in general media, \cite{obukhov1}. Therefore it is necessary to extend the analysis and study more complex media, for instance with anisotropic or magneto-electric properties. In any case, our analysis, along with those in ref. \cite{obukhov1,pfeifer,penfieldhaus1,mikura}, shows that the Abraham choice of the ``correct'' momentum of a light pulse is only one possibility, simple and useful for the description of isotropic media, but not at all an unique one.

\appendix

\section{Electromagnetic plane wave solution in a linear, homogeneous, 
isotropic, moving medium}\label{appa}
The macroscopic Maxwell equations and the constitutive relation for a \textit{linear} medium are given in covariant form by \cite{obukhov1}
\begin{align}
\partial_{\nu}H^{\mu\nu}={}&J^{\mu}_{\rm ext},\label{inhomo}\\
\partial_{\mu}F_{\nu\rho}+\partial_{\nu}F_{\rho\mu}+\partial_{\rho}F_{\mu\nu}={}&0,\label{homo}\\
H^{\mu\nu}={}&\frac{1}{2}\chi^{\mu\nu\rho\sigma}F_{\rho\sigma}.\label{constitutive}
\end{align}
In particular, we consider an \emph{homogeneous}, \emph{isotropic}, \emph{non-dissipative} and \emph{non-dispersive} medium in motion and therefore we can express the constitutive tensor $\chi^{\mu\nu\rho\sigma}$ in terms of the well-known Gordon optical metric $\gamma^{\mu\nu}$ by
\begin{align}
\chi^{\mu\nu\rho\sigma}:=\frac{1}{\mu_0\mu}\left(\gamma^{\mu\rho}\gamma^{\nu\sigma}-\gamma^{\mu\sigma}\gamma^{\nu\rho}\right),\label{isotropic}
\end{align}
where,
\begin{align}
\gamma^{\mu\nu}:= g^{\mu\nu}+\frac{(n^2-1)}{c^2}u^{\mu}u^{\nu},\label{gordonmetric}
\end{align}
as it was first derived by Gordon in 1923 \cite{wgordon}. The Minkowski metric $g_{\mu\nu}$ is defined as $g_{\mu\nu}={\rm diag}(c^2,-1,-1,-1)$, $\varepsilon$ is the dielectric constant, $\mu$ the relative permeability, $n:=\sqrt{\varepsilon\mu}$ the refractive index, $u^{\mu}$ the $4$-velocity of the moving medium and $c:=1/\sqrt{\mu_0\varepsilon_0}$ the velocity of light in vacuum. 
If we look for plane wave solutions of the form
\begin{equation}\label{ansatzpw}
F_{\mu\nu}=\Re\left\lbrace\tilde{F}_{\mu\nu}e^{ik_{\lambda}x^{\lambda}}\right\rbrace,
\end{equation}
then the optical metric $\gamma^{\mu\nu}$ determines the dispersion relation:
\begin{align}
\gamma^{\mu\nu}k_{\mu}k_{\nu}=0.\label{dispersionrel}
\end{align}
We consider the one-dimensional slab problem for which $k_{\mu}$ and $u^{\mu}$ are of the form
\begin{align}
u^{\mu}=(\gamma, \gamma v,0,0),\qquad
k_{\mu}=\left(\omega,-k,0,0\right),\label{uuu}
\end{align}
where $v$ is the velocity of the slab and $k$ the wave number. Using (\ref{uuu}a) in (\ref{gordonmetric}), the explicit expression for $\gamma^{\mu\nu}$ reads,
\begin{equation}
\gamma^{\mu\nu}=\left( \begin{array}{cccc}
[1+(n^2-1)\gamma^2]/c^2 &(n^2-1)\gamma^2\beta/c &0&0 \\
(n^2-1)\gamma^2\beta/c &-1+(n^2-1)\gamma^2\beta^2 &0&0 \\
0&0&-1&0 \\
0&0&0&-1 \\
\end{array}\right).\label{gammaaa}
\end{equation}
Then, using (\ref{uuu}b) and (\ref{gammaaa}) in (\ref{dispersionrel}) and solving for 
$\omega$ in terms of $\beta$, $n$ and $k$ we obtain the dispersion relation of light inside the 
moving medium, $\omega(k)=v_{\beta}k$, where $v_{\beta}$ is the phase velocity of the waves inside the moving medium, 
given by (\ref{phasevelocity}).

Finally, inserting the ansatz (\ref{ansatzpw}) into the Maxwell equations (\ref{inhomo}) and 
(\ref{homo}), and using (\ref{isotropic}), (\ref{gordonmetric}), one can find, after some 
algebra, general explicit expressions for the field strengths $\bm{E}(x,t)$ and 
$\bm{B}(x,t)$. However, here we just display the properties that are used in the main text, 
namely $\hat{\bm{x}}\cdot\bm{E}(x,t)=0$ and (\ref{bfield}). 
Relations (\ref{phasevelocity}) and (\ref{bfield}) can be also obtained by applying the appropriate relativistic transformation of the fields from the rest frame of the medium. Notice however that the relativistic transformation of the phase velocity is in general different from the usual particle velocity transformation, but they do coincide if the phase velocity is parallel to relative velocity. See, for instance, \cite{bachman} and \cite{zhang}.

\bibliographystyle{elsarticle-num}

\end{document}